\documentclass[aps,10pt,showkeys,showpacs]{revtex4}
\pdfoutput=1
\usepackage{amssymb}
\usepackage{latexsym}
\usepackage{epsfig}

\usepackage[utf8]{inputenc}
\usepackage{amsfonts}
\usepackage{amsmath}
\usepackage{amsthm}
\usepackage{amscd}
\usepackage{epsfig}
\usepackage{epstopdf}
\usepackage{graphicx}
  \graphicspath{{/}{figures/}}
  \DeclareGraphicsExtensions{.pdf,.png}
\usepackage{comment}
\usepackage{enumitem}

\begin{document}

\title{ GARCH(1,1) model of the financial market with the Minkowski metric}

\author{Richard Pincak  $^{1,2}$\footnote{pincak@saske.sk}, Kabin Kanjamapornkul $^{3}$\footnote{kabinsky@hotmail.com}}
 \affiliation{ $^{1}$
Institute of Experimental Physics, Slovak Academy of Sciences,
Watsonova 47,043 53 Kosice, Slovak Republic.\\$^{2}$ Bogoliubov Laboratory of Theoretical Physics, Joint
Institute for Nuclear Research, 141980 Dubna, Moscow region, Russia\\ $^{3}$	 Department of Survey Engineering.\\ Faculty of Engineering, Chulalongkorn University. }

\begin{abstract}


We  solved a stylized fact on a long memory process of volatility cluster phenomena by
 using Minkowski metric for GARCH(1,1) under assumption that price and time can not be separated. We provide a  Yang-Mills equation in financial market and anomaly on superspace of time series data as a consequence of the proof from the general relativity theory.
We used an original idea in Minkowski spacetime embedded in Kolmogorov space in time series data with behavior of traders.
 The result of this work is equivalent to the dark volatility or the hidden risk fear field induced by the interaction of the behavior of the trader in the financial market panic when the market crashed.

\pacs{98.80.-k, 04.50.Gh, 11.25.Yb, 98.80.Qc}
\keywords{GARCH; Time series; Minkowski metric; Volatility; Yang-Mills; Market panic}

 \end{abstract}
 \date{\today}

\maketitle

\section{Introduction}\label{sec:intro}

Bacheiler \cite{normal}  was the pioneer of the
efficient market hypothesis (EMH) \cite{fama} established
on an assumption that the price obeys the random walk \cite{levy} in a spectacular
market \cite{manibrot}. In economics, the time series data is a record in an Euclidean plane where price and time are independent to each other. In the general relativity, we cannot separate price and time under the Euclidean space with the $T_{2}-$ Hausdorff separation \cite{haus}. It is still a hidden loop space in time series data where price and time cannot be separated in the analogy with the Minkowski spacetime. There exists an empirical analysis and it is found stabilized fact on volatility cluster inducing long memory process in the financial time series.
  B. Mandelbrot introduced a new model of GARCH,
 so-called fractional Brownian motion and gave a result in FGARCH.
 The main problem in FGARCH is an infinite variance and the stylized fact still happens in the form of the power law distribution.
With the development of the quantum field theory \cite{atiya}, the Chern-Simons current induced from the interaction of the dark energy in a living organism \cite{current} , loop gravity \cite{loop1}
and some new emerging disciplines such as econophysics \cite{econophysics},
complex system  \cite{cohesive} and behavior economic theory \cite{cluster2}, economists and econophysicists  gradually
get deeper understanding on the market hidden geometry induced from the interaction of buying and selling orderbook submissions of traders  and
hidden behavior  \cite{ssm} of traders in financial market \cite{anomaly} which we empirical observed from the financial time series data by using GARCH model \cite{garch2}.
Some experts in the complex system and fractal geometry \cite{cluster} had been argued that EMH cannot be
use to explain some weird phenomena of financial time
series, e. g., volatility clustering phenomena. Scientists realized that the Euclidean plane is not enough to explain
such a chaotic phenomenon without adding something more of the statistical theory of spinor field \cite{pauli} in  Wilson loop   phase transition  \cite{phase2} to the plane.
We can extended the Euclidean space to a non-Euclidean space under the superspace in the time series data with the supersymmetric property. In the Yang-Mills theory \cite{yang}, the superspace is  composed of a quantum foam and  of the lattice gauge field with some extra properties of the Khovanov cohomology  \cite{khovanov} and the Chern-Simons theory \cite{chern}.

 The problem comes from the classical
financial mathematics. We just want to perform fundamental
investigations to find an adequate stochastic processes matching
the real financial time series data \cite{cohomo} without considering the market microstructure with the underlying orderbook \cite{orderbook} of the financial market scaling \cite{memory} and the hidden structure of the financial market as a D-brane \cite{brane} with extradimensions \cite{pincak3}.
 The financial time series
data in the moment of the financial market crashed \cite{crash} are typical experimental results which we have for the complex
systems \cite{crash2}.
In the analysis of an empirical work of the financial
time series, nonlinearities and non-stationarities  \cite{emd} are shown with the stylized fact of a volatility cluster
phenomena in a long memory effect.
 Some statisticians  and   economists typically predicted macroeconomics of the time
series  and the financial time series  by using statistical data analysis of, ARIMA, ARCH, GARCH and the Markov switching model using the assumption of the linearity and the stationary time series process which cannot be found in the most of the typical financial time series data.
The main defect of the ordinary least square (OLS) \cite{kolmogorov} ARIMA and GARCH model is based on fitting the problem
with the parameters of fitting or learning with a single stochastic process
for infinite factors which is governed by an infinite stochastic
process influence on the future expectation price.
When we add one point of the future price and fit curve by using the data
mining tool for the regression with GARCH(1,1), the coefficient
of the equation which we used to describe the historical
data will update and change the historical path so that
it makes a non-realistic situation.
In real life, we cannot change
a historical event but we can influence an expected future event. We call this problem a prior effect in the scaling behavior of the time series data.
There appears a prior effect on the wavelet transformation when we
add a one more point and do a wavelet transform again, so
the result is not the same as the previous result.
In the case of Fourier transform or a Fourier series of an approximate function by using the periodic function, we called this the Joseph effect.
In the Hilbert-Huang transform, we call this the end effect, in econometric we also
can notice this by an ordinary least square (OLS) method so that
beta was changed when you add more point.
  This problem will also happen in the hidden Markov model or the Bayesian network
because of all the families of the tool are based on the simple statistic, not on the superstatistics or the hyperstatistical theory in which it is contained an
infinite stochastic process
for fitting.
 We call this effect in a 
general term as a prior effect or a prior problem, the problem
that you lie people on by just changing your historical
record in the event of what you did.
In the present time, physicists  \cite{pincak1} defined another type of
the geometrical object on the financial market, which may live on the
string \cite{pincak2} and the D2-brane world \cite{pincak5}.
 In order to build a new mathematical superstructure
suitable for the financial market we need to take into account also a
psychology factor of traders in the model of the equation.
The spinor field in the model cannot be the price of the direct buy, it should
be a covering space of the price arising from the interaction of the trader
with the different expectations on the supply and the demand of the stock.

 It was found that the state space of the financial
market is considered as a smooth manifold containing events as a
fiber space endowed with the particular affine connection.
Let the behavior of the trader be a sequence of the state spaces of the Riemannian
manifold, so-called be behavior state space model, a cohomological
 sequence of agents and dual agents.
The representation of the space trader is a so-called agent $A_{t}$
at time $t$ and the sequence of the expectation price of the trader is a so
called dual agent $A^{\ast}_{t} $, the sequence is a homological
algebra which induces a cohomological sequence as an
evolution feedback of price evolution.
Let us consider the path connected component of the quotient group between the circle and the boundary defined by the feedback evolution between the shape of
the endpoint of time series of predictor and predictant after
performing the transform into zeros band the transformation of the empirical mode decomposition.

 In this work, we use  a new approach to solve this problem by using differential geometry.
 We build a new GARCH(1,1) model based on  \cite{arch1}, \cite{risk} in Wilson loop \cite{wilson} of the connection  for the description of the behavior of the interaction of the 
traders in the financial market. The gauge group action with the connection in the Chern-Simons current is a source of the spinor field appearing as a hidden risk fear field in the financial market.

The paper is organized as follows: in section II,
 we review GARCH(1,1) invented by Bollerslev in another framework of differential
geometry. The allowed volatility should be minus in order to interpret the result of the risk field.
In order to open a  new door to solve the stylized fact, we use a Minkowski space instead of the single Kolmogorov system for the volatility cluster. Volatility is used to express an emotional trade by defining a fear factor in the stock
market or the stress factor.
 We discuss on stylized fact and give a new definition of the Grothendieck and the De Rham cohomology for the financial market with the Minkowski metric for GARCH(1,1).
In section III, we provide a proof of volatility clustering phenomena in the hyperbolic space with the Yang-Mills equation for the financial market.
In section IV, we summarize the results of work and discuss future work.

\section{GARCH(1,1) and Stylized Facts}

\newtheorem{Definition}{Definition}
\newtheorem{Theorem}{Theorem}

 Let $\sigma_{t}(\mathbf{r}_{t})^{2} $ be a volatility of the stock return. Using this mathematical transformation, we induce a dark volatility equivalent to a
dark matter in the theoretical physics discovered by Wolfgang Pauli. The experiment which took place more than 20 years after
death of Pauli confirmed neutrino in the balance of energy and momentum equation in his quantum field equation.
Let $F$ be a fiber bundle.
Let $g_{ij}\in G$ be a cycle and $g^{ij}\in G^{\ast}$ be a cocycle, a group action acts on the fibre bundle of manifold.
We have $g_{ij}=\sqrt{g_{ij}}\sqrt{g_{ij}}$.
Let $x_{t}$ be a time series in the section of the Riemannian manifold, $x_{t}\in T_{x_{0}}X=p^{-1}(X).$
So,  we have a differential 2-form of time series arising from the scalar production of the group transformation of the vector bundle, so-called principle bundle of time series,
$\sigma_{t}^{2}=<x_{t},x_{t}>=g_{ij}||x_{t}||^{2} $
 with $g_{ij}=1 $.  The volume form is invariant with the respect to translation and rotation and the volatility in the normal form
will preserve the transformation.
In this paper, we are interested in the case of the dark volatility, $\sigma_{t}^{2}<0$, it happens when
we change the Euclidean metric to the Minkowski metric $g_{ij}=-1$ that allows the volatility minus, $\sigma_{t}^{2}<0$, because
\begin{equation}\sigma_{t}^{2}=<x_{t}.x_{t}>=g_{ij}||x_{t}||^{2}=-||x_{t}||^{2}.
\end{equation}

 In this case, we have an assumption that every risk from the stock investment in the stock market induces a fear field or risk field, so called volatility network. We denote this field $C_{\ast}(\sigma_{t,i}^{2}),i=1,2,3,\cdots ,n$ for $n-$ stock in stock market.
Let us denote a price in stock market as $p_{t}(s_{i})$, where $s_{i}$ is a code of stock $i.$
We have an assumption that may be a source of risk field coming from the trader willing to win a high profit from stock market without their excess demand, but on the other side, they use too much leverage. This situation will induce both covariance and contravariance tensor field of the volatility of the price. Let us consider the Jacobian flow of GARCH(1,1) equation. Recall equation $(2)$ of GARCH(1,1).

Discrete time volatility models were developed before the high-frequency data became readily available, and are typically applied to the daily or the lower frequency returns. Most discrete time models for the daily financial return $r_{t}$ satisfy the canonical product structure
\begin{equation}
r_{t}=\sigma_{t}\epsilon_{t}
\end{equation}
returns with multiplicative shock, $\epsilon_{n}$.

Here the observed financial return $r_n$ is modeled as the product of an iid innovation $\epsilon_t$
and a positive scale factor  $\sigma_{t}>0$. One usually assumes that
$\epsilon_t$ has a mean value zero and, for the
standardization, a unit variance. Specific models differ in their specification of the scale
factors; an example is the stationary GARCH(1,1) recursion
\begin{equation}
\sigma_{t}^{2}=\kappa+\alpha r_{t-1}^{2}+\beta\sigma^{2}_{t-1}
\end{equation}
where $\kappa,\beta,\alpha >0$ and $\alpha+\beta <1.$

The scale factors $\sigma_{t}$ (Jacobian in this model) are not observed (hidden), and one may use the daily close-to-close
returns $r_n$ to estimate and evaluate the models for $\sigma_{t}$. Let us consider GARCH(1,1) for the series of returns $r'_{t}=log(r_{t})$ which can be written as an additive shock,
\begin{equation}
r'_{t}=\delta+\epsilon_{t} =\delta +\eta_{t}\sqrt{h_{t}},\nonumber
\end{equation}
and
\begin{equation}
h_{t}=\alpha_{0}+\alpha_{1}+\epsilon_{t-1}^{2}+\beta h_{t-1}.
\end{equation}

The main problem of GARCH(1,1) is that the return is not in the log scale but we have a constant Euclidean distance. In this work, we change the metric to the hyperbolic distance in order to solve the volatility cluster phenomena. We found a complete solution in the complex time scale with the conjugate solution in the minus time scale.   We called the volatility in the inert hidden time scale a hidden fear field or dark volatility
${\sigma_{t}^{\ast}}^{2}(A_{\mu})$ with a hidden variable of a risk field $A_{\mu}$ as a connection in the gauge field theory.

The difference between the volatility and the dark volatility are the classes of  different topological spaces as their mathematical objects.

The volatility $\sigma_{t}^{2}$ is based on the statistical theory under the normal distribution and the usual treat is a random variable satisfying the Kolmogorov axiom of probability under the Hausdorff space with an outer spinor field.
We define the dark volatility to be the general form of the normal distribution over the spinor field in the quotient superspace
 in the time series data under the $T_{0}-$ separation axiom. The volatility ${\sigma^{\ast}_{t}}^{2}(A_{\mu})$ is a spinor field of the behavior of the traders with a group action by the market cocycle $g$. The example of the dark volatility
is a left chiral projection into the
Euclidean plane with the left translation of the mean value only. It is just a volatility in the normal distribution if we take a group action as a left translation in the mean value. In the equilibrium of the statistical system, we have a normal distribution with the identity of the left translation of the gauge group appearing as a Gaussian random  variable $\sigma_{t}^{2}:=0$, the identity of the Lie algebras as a result of the left adjoint map as volatility in the Euclidean plane.
 If we extend the group action as a spinor field action to the time series in the form of a gauge potential field $x_{t}:=gA_{\mu}$ translation in the parallel direction of the left and the right translation with an extra property of the spinor invariant, we will get a new definition of the volatility as the Jacobian flow in both left and right direction simultaneously in the superspace of the time series data.

 In the Minkowski space of time series data, we get supply and demand as focus points with the reflection of price in the light cone as a mirror symmetry from the left to the right chiral state in the financial market with the supply and the demand side.
Therefore, we define the volatility in GARCH(1,1) as a gauge group action in the Jacobian flow with $g:=g_{ij}=\sigma_{t}^{2}$ or $h_{t}$ term.
We substitute $h_{t}$  with $g_{ij}(t)$,  so we have a Jacobian flow equation of the parallel transport GARCH(1,1) model
\begin{equation}g_{ij}(t)=\alpha_{0}+\alpha_{1}||\epsilon_{t}||^{2}+\beta g_{ij}(t-1),
\end{equation}
so we have an extension of the Euclidean plane to the non-Euclidean plane with the same  as GARCH(1,1), so called Minkowski GARCH(1,1) in this context,
 \begin{equation}\alpha_{1}||\epsilon_{t}||^{2}=   g_{ij}(t)-\alpha_{0}-\beta g_{ij}(t-1), \end{equation}
where $g_{ik}$ is an inverse of $g^{kl}$ representation for the market risk cycle and risk cocycle.
In economic, most people assume $||\epsilon_{t}||^{2}\sim N(\mu,\sigma^{2})$ to be a random variable with random walk.
In this work, we assume that this term is an induced spinor field from the interaction of the behavior of the trader with a risk aversion with no concept of the normal distribution. We use the differential geometry and the gauge theory approach for the approximation of this term with the Laurent polynomial in the Chern-Simons theory in the Kolmogorov space in the time series data with $x\in X,y\in Y,x_{\ast}\in X^{\ast}, y^{\ast}\in Y^{\ast}$. We have
\[||\epsilon_{t}(A_{\mu})||^{2}=<\frac{x\wedge x^{\ast}}{C},\frac{y\wedge y^{\ast}}{C}>   , x^{\ast} (x\otimes y)=C\]
where $C$ is a constant value representing a memory effect in the volatility clustering phenomena of the time series data and $A_{\mu}$ is a connection.

Let $ p:  \Omega^{\ast}(\wedge ^{2} T_{x_{0}}\mathbb{R}\otimes \mathbb{R})\rightarrow \mathbb{R}$ with section matrix be a half volatility model, i.e. the matrix of error of the time series prediction $\epsilon_{t}$ in the section of the supermanifold of the financial market, $p^{-1}(\mathbb{R})=[\epsilon_{t}]$.

It is independent of the coordinate system, so we can use the Jacobian flow of the coordinate transform across a stack of time scale going in both directions of future and past,
\begin{equation}d\epsilon_{t+1}=\sqrt{g_{ij}(t)} d\epsilon_{t},\hspace{1cm}d\epsilon_{t-1}^{\ast}=\sqrt{g^{ij}(t)} d\epsilon_{t}^{\ast}
\end{equation}

so we have a Jacobian flow as Minkowski GARCH(1,1)
model,
\begin{equation}
|| \epsilon_{t}||^{2} =a+b \frac{dg_{ij}(t)}{dt}
\end{equation}

where $a, b$ are constants and $g_{ij}(t)$ is a market cocycle.
A parallel transport is a Wilson loop of connection  $W_{\alpha,\beta}(A_{\mu})$ over a principle fibre of financial supermanifold as arbitrage
opportunity in the financial market cocycle $(\alpha,\beta)$. It is obtained from the Jacobian flow of the market cocycle over the fibre space by
\begin{equation}
A_{\mu}=\Gamma_{ij}^{\mu}=\frac{1}{2}g^{\mu l}(\partial_{j}g_{il} +\partial_{i}g_{lj}-\partial_{l}g_{ji}).
\end{equation}
The intepretation of $A_{\mu}$ is an error of the expected price from the physiology of the time series data, while trading induced from the interaction of the risk aversion and the hidden risk fear field from the behavior of the fundamentalist $A_{\mu=1}=A_{1}$. It is a chartlist or noise trader $A_{\mu=2}$ and bias  traders $A_{\mu=3}$ in the financial market in the positive and the negative forward looking to the future price of the market as a group operation of the gauge group of an expected market cocycle $g$ to the gauge field $A_{\mu}$.

Here $A_{\mu}:=\Gamma_{ij}^{\mu}$ is a risk gauge potential induced form market cocycle $g=g_{ijk}=g_{ij}(k)$ as gauge group action
\begin{equation}
[g,A_{\mu}]=\sigma^{2}=A_{\mu}g-gA_{\mu}=Ad_{g}\{A_{\mu}\}.
\end{equation}

We have $gA=Ag+||\epsilon||^{2}$ with the gauge group action $A_{\mu}^{g}=g^{-1}A_{\mu}g  +g^{-1}\partial_{i}g$.
The Wilson loop of the behavior trader in the connection $A_{\mu}$ of fibre space is defined by
\begin{equation}
W_{\alpha,\beta}(A^{g})=W_{\alpha,\beta}(A)-\frac{1}{24\pi^{2}}\int d^{3}x \epsilon^{ijk} <g^{-1}\partial_{i}gg^{-1}\partial_{j}gg^{-1}\partial_{k}g>.
\end{equation}
From the typical classical differential geometry, the equation above allows us to work over the connection of the fibre space as an extradimension in the moduli space of Killing vector field approach.
We just use the tool for the computation of the expected state in our new definition of the spinor field in the time series data. The spinor field induces coupling state between 4-types of tensor fields over Killing equation above.
\begin{Definition}
Let support spinor be an arbitrage opportunity ,$\Gamma_{ij}^{m}$ in the financial market over moduli state space model in the time series data as a connection preserving scalar product over the parallel translation along the fibre space of the physiology layer in the time series data.
\end{Definition}

We define a Chern-Simons current as an arbitrage opportunity density by,
\begin{equation}
 J_{A}^{\mu}=\partial \mu F^{\mu\nu} +[A_{\mu},F^{\mu\nu}.]
\end{equation}
\begin{equation}
\frac{1}{4}<\ast F^{\mu\nu} F_{\mu\nu}>=\partial K^{\mu}
\end{equation}
where
\begin{equation}
K^{\mu}=\epsilon^{\mu\alpha\beta\gamma} <\frac{1}{2} A_{\alpha} \partial_{\beta} A_{\gamma} +\frac{1}{3}
A_{\alpha} A_{\beta} A_{\gamma}>
\end{equation}
where a connection $\Gamma_{\alpha\nu}^{\mu}=(A_{\alpha})_{\nu}^{\mu}$, where $A$ is a Chern-Simons current in the financial market.
An arbitrage or arbitron is  a Chern-Simons anomaly, $K^{\mu}$ or an anomaly current induced from the behavior of the trader as a twist ghost field between 2 sides of the market. This field of the financial market induces a Ricci curvature in the financial market as an arbitrage opportunity for each connection in the coupling ghost field of the behavior of the trader. This curvature blends the Euclidean plane of the space of the observation to twist with the mirror plane behind the Euclidean plane and wraps to each other as the D-brane and the anti-D-brane interaction of the superspace of the financial time series data.
\begin{equation}
F_{\mu\nu}=R^{\mu}_{\nu\alpha\beta}(\Gamma) = e_{\alpha}^{\mu} R_{b\alpha\beta}^{a} e_{\nu}^{b}.
\end{equation}

\begin{Definition}
The Ricci tensor in the time series data is a contraction of the curvature tensor of the risk defined by
$R_{ik}(\sigma_{t}^{2})=R_{ikl}^{j}(\sigma_{t}^{2})$ with the respect to the natural frame of the connection
\begin{equation}
F_{\mu\nu}:=R_{ik}(\sigma_{t}^{2})=\partial_{k}\Gamma_{ji}^{j}- \partial_{j}\Gamma_{ki}^{j}+\Gamma_{km}^{j}\Gamma_{ji}^{m}-
\Gamma_{jm}^{j}\Gamma_{ki}^{m},
\end{equation}
where  $Ad_{g}\{A\}:=[g,A]:=g\times \Gamma_{ji}^{j}:=g\times A= \sigma_{t}^{2}.$
\end{Definition}

The market equilibrium occurs when the Ricci curvature is zero, $R_{ik}=0$ (arbitrage opportunity disappears and the physiology of the time series data contains no curvature).

The solution of the Patterson-Godazzi differential equation is a curvature of the
superspace in the time series data.

If $\partial \sigma_{t}^{2}=1$ or constant, we have an equal risk field in the market or the risk free market $\sigma_{t_2}^{2}-\sigma_{t_1}^{2}=1$, so we have the price persistence with the shock given by
\begin{equation}\frac{\partial p_{t}}{\partial \sigma_{t}^{2}} =p_{t_2}-p_{t_1}>0,\end{equation}
but in the case of the dark volatility (as a hidden fear field) that induces
an inertia in the price in hidden direction (and also if 
$\partial \sigma_{t}^{2}=-1$),
\begin{equation}\frac{\partial p_{t}}{\partial \sigma_{t}^{2}} =p_{t_2}-p_{t_1}<0.\end{equation}
We conclude that $\sigma_{t}^{2}\in \mathbb{H}$ is called the dark volatility. We can monitor or measure on the border of the universe outside of our galaxy with the light speed only.

Let us define an underlying coordinate of the Jacobian flow. If $g_{ij}=1$, we use a $S^{2}$ Riemann sphere as  GARCH(1,1) coordinate. When
 $-1<g_{ij}<1$, we use a cone or hyperbolic coordinate and when $g_{ij}=-1$, we use the Minkowski space.

\subsection{Dark Volatility}

In this work, we assume that the financial market be is an asymmetric superspace $X_{t}/Y_{t}$ with the left and right supersymmetry in the Yang-Mills theory. The left chiral supersymmetry is a source of the demand $D\in X_{t}$ and the right chiral supersymmetry is a source of the supply in the hidden space $S\in Y.$ The duality in the general equilibrium with the invisible hand is a source of the hidden dual superspace with the hidden demand $D^{\ast}\in X^{\ast}$ and the hidden supply $S^{\ast}\in Y^{\ast}$. Most economists treat the price $p_{t}\in \mathbb{R}^{n}$ as the time series data embedded in the Euclidean plane with an outer extra property of the spinor field. We use an alternative approach of the differential geometry.
Our price is defined by an extension of the Euclidean plane to a quaternionic projective space $\mathbb{H}P^{1}$,
$p_{t}(S,S^{\ast},D,D^{\ast})$ is this space with an extraproperty of the spinor field from the interaction of the behavior of the trader from the supply and demand side as the connection $A_{\mu}$ in the fibre space with the Hopf fibration.
In econophysics, market depth is defined by a Taylor expansion of the price over an excess demand $D$.
We have an equivalent class of the supply and demand when the price is a nonequilibrium of the matching by
\begin{equation}
S^{\ast}\sim D\leftrightarrow S^{\ast}\alpha= D, S\sim D^{\ast}\leftrightarrow S\beta= D^{\ast},
 \end{equation}
where $(\alpha,\beta)$ is an equivalent class market cocycle of the price matching.

  We assume that the risk is composed of the observation risk and the hidden risk in the supersymmetric property of the left and the right chiral supersymmetry. The left chiral risk is the volatility in economics with the systematic risk $\sigma^{2}_{t}$ and the right chiral risk is the dark volatility, the market risk with the hidden risk fear field induced from the behavior of the trader $\sigma_{t}^{2}(A_{\mu=1,2,3})$. Let price be a transition state of the coupling hidden state between the supply pair, $(S,S^{\ast})$ and demand pair $(D,D^{\ast})$.

 Hence, we have 2 systems of a hyperbola with the focus point $D=\beta S^{\ast},D^{\ast}=\alpha S$ in the hidden supply and the hidden demand and the center in the demand and supply with pairs states twists space to each other with 2 Minkowski cones with the superdistribution $pp(k)$ of the price momentum $k$.
If we rotate the hyperbola light cone with $\theta=\frac{\pi}{2}$, we get 2 systems of equations,
\begin{equation}
  Cone(x,y,pp(k)) :=\{(x,y)| (x-D)^{2}-(y-S)^{2}=k\} \nonumber
\end{equation}
\begin{equation}
 Cone(x^{\ast},y^{\ast},pp(k^{\ast})) :\{ (x^{\ast},y^{\ast})|(x^{\ast}-D^{\ast})^{2}-(y^{\ast}-S^{\ast})^{2}=k^{\ast}\}
 \end{equation}

A Wilson loop of the time series data, denoted as $W_{\alpha,\beta}(A_{\mu})$ is a twistor in acomplex projective space $\mathbb{H}P^{1}$. It turns one side of the Euclidean plane twist into another side of the plane with the help of modified Mobius map $z=\frac{1}{z}$.

We have $W_{\alpha,\beta}(\frac{\lambda S}{\lambda D})=\frac{D}{S}$.
The equivalent class of the supply and demand is induced from the queuing process of $0$ to $\infty$ by using modified Mobius map $z=\frac{1}{z}$. Each of the components is defined by gluing an equivalent class of the supply and demand $S\in [s],D\in [d]$ such that $[s] \sim [d]$ if and only if there exists $\lambda_{i}$ such that $x_{i}=\lambda_{i} y_{i},\lambda=(\alpha,\beta)$.

The projective coordinate is a local coordinate on  the Minkowski space, a Wilson loop of the unitary operator in the quantum mechanics. Let $g_{ij}$ be Jacobian matrix of the supply and demand, we have a shallow to the hidden state of the transformation by using the Wigner ray transform
\begin{equation}
g_{ij}\mapsto \lambda g_{ij},\lambda \in \mathbb{U}(1)\simeq Spin(2)
\end{equation}
where $\lambda =\sigma_{x},\sigma_{y}$ are pauli matrix in $Spin(2)$ group.

\begin{Definition}
Let $S$ be a supplied linear compact operator in Banach space and $D$ be demand operator.
A Wilson loop in the time series data at a general equilibrium point in the financial time series data is a point of price $x_{t}(S,D)$ such that there exists a ray of the unitary operator $\lambda \in  SU(2)\simeq Spin(3) $,
\begin{equation}
W_{\alpha,\beta}(<S,D>):= <\lambda S,\lambda D>=\lambda<S,D>=<S,D>
\end{equation}
\end{Definition}
 The determinant of the correlation matrix or a wedge product of the column of the correlation of all returns of stocks in the stock market can be minus one and induces a complex structure as a spin structure in the principle bundle of the time series data. We found that when
$det(Corr(M))^{2}=1$, where $Corr(M),M\in SL(2,\mathbb{C})$ is a correlation matrix of the financial network $M$ with Mobius map. This space of the time series data is in an equilibrium with 2 equilibrium points.

One is a real and explicit form, the other is hidden and in the complex structure like dark matter or invisible hand in the economics concept. If $det(Corr(M))=0$, 
the space of time series data is out of equilibrium and it can induce the market crash with a more herding behavior of the noise trader.
This result is related to the definition of the log return of an arbitrage opportunity in econophysics.

The section of the tangent of the manifold is a Killing vector field of the manifold of the financial market. We introduce 3 types of the Killing vector field for the market potential field of the behavior of the traders in the market.
\begin{equation}
A_{\pm,1}(f)=\frac{\partial}{\partial f}:SO(2)\rightarrow Spin(2):M\rightarrow T_{x}M
\end{equation}
\begin{equation}
A_{\pm,2}(\sigma)=\frac{\partial}{\partial \sigma}:SO(2)\rightarrow Spin(2):M\rightarrow T_{x}M
\end{equation}
\begin{equation}
A_{\pm,3}( \omega)=\frac{\partial}{\partial \omega}:SO(2)\rightarrow Spin(2):M\rightarrow T_{x}M
\end{equation}
where $+$ means optimistic behavior of the trader, $-$ means the pessimistic trader, $f$ is a fundamentalist trader, $\sigma$ is a noise trader and $\omega$ is a bias trader. $A_{1}$ is an agent field of the adult behavior trader as fundamentalist. $A_{2}$ is an agent field of teenager behavior trader as herding behavior or the noise trader. $A_{3}$ is an agent field of the child behavior of the trader or the bias trader. The behavior is spin up and down as the expected states in the market communication layer of the transactional analysis framework. The upstate is signified as the optimistic market expected state and pessimistic as the market crash state of an intuition state of the forward looking trader cut each other as in the transactional analysis framework of the market communication between the behavior of the expected state and the behavior of the market state.

The strategy of the fundamentalist (we denoted shortly by $f$) is the expected price in a period, where the price is in the minimum point $s_{4}$ and maximum point $s_{2}$ (crash and bubble price). Then the optimistic fundamentalist, $f_{+}$ will buy at the minimum point $s_{4}$ of the price below the fundamental value and the pessimistic fundamentalist $f_{-}$ will sell the product or short position at the maximum point of the price if the maximum point is over a fundamental value. The strategy of chartist or noise trader, $\sigma_{\pm}$ is different from the fundamentalist. Let $\omega$ be a basic instinct bias behavior of trader.

We can use the Pauli matrix $\sigma_{i}(t)$ to define a strategy of all agents in the stock market as the basis of the quaternionic field span by the noise trader and the fundamentalist. Let $+1$ stay for buy at once and $-1$ for be sell suddenly. Let $+i$ stay for being inert to buy and $-i$ for being inert to sell.
The row of the Pauli matrix represents the position of the predictor and the position in the column represents the predicted state.
\begin{Definition}

For the noise trader, we use an finite state machine for the physiology of the time series to the accepted pattern defined by
\begin{equation}
A_{2}:=\sigma(t)= \sigma_{y} =  \left[   \begin{array}{ccc}
 &s_{2}{\ast}(t) &s_{4}^{\ast}(t)\\
s_{2}(t)&0&-i\\
s_{4}(t)&i&0\\
\end{array}\right ]=  \left[   \begin{array}{ccc}
 &s_{2}{\ast}(t) &s_{4}^{\ast}(t)\\
s_{2}(t)&0&\sigma_{+}\\
s_{4}(t)&\sigma_{-}&0\\
\end{array}\right ]
\end{equation}
where $\sigma_{y}$ is a Pauli spin matrix.
\end{Definition}

For the fundamentalist trader, we use an finite state machine for the physiology of the time series to the accepted pattern defined by

\begin{equation}
 A_{1}:=f_{\pm}=-W(\sigma_{z}) =  \left[   \begin{array}{ccc}
 &s_{2}^{\ast} &s_{4}^{\ast}\\
s_{2}(t)&0&-1\\
s_{4}(t)&1&0\\
\end{array}\right ]=\left[   \begin{array}{ccc}
 &s_{2}^{\ast} &s_{4}^{\ast}\\
s_{2}(t)&0&f_{-}\\
s_{4}(t)&f_{+}&0\\
\end{array}\right ]
\end{equation}

Let $\omega$ be a biased behavior $A_{3}$ of the market micropotential field.
Since $[\sigma_{y},\sigma_{z}]=2i\sigma_{x}$,

 \begin{equation}
[f,-W^{-1}(\sigma)]=2i \omega
\end{equation}
where $W$ is Wilson loop or knot state between the predictor and the predictant for the time series data and $W^{-1}$ is the inverse of the Wilson loop for the time series data (unknot state between the predictor and the predictant).

We have an entanglement state inducing the strategy of the noise trader as herding behavior explained by
\begin{equation}
s_{2} \stackrel{-i}{\rightarrow } s_{4}^{\ast}\hspace{0.5cm}, s_{4} \stackrel{i}{\rightarrow } s_{2}^{\ast}
\end{equation}
with
\begin{equation}
W_{\alpha,\beta}( \left[   \begin{array}{c}
 s_{2}  \\
s_{4}^{\ast} \\
\end{array}\right ] )= \left[   \begin{array}{c}
 s_{4}^{\ast}  \\
s_{2} \\
\end{array}\right ]
\end{equation}

\begin{figure}[!t]
\centering
\epsfig{file=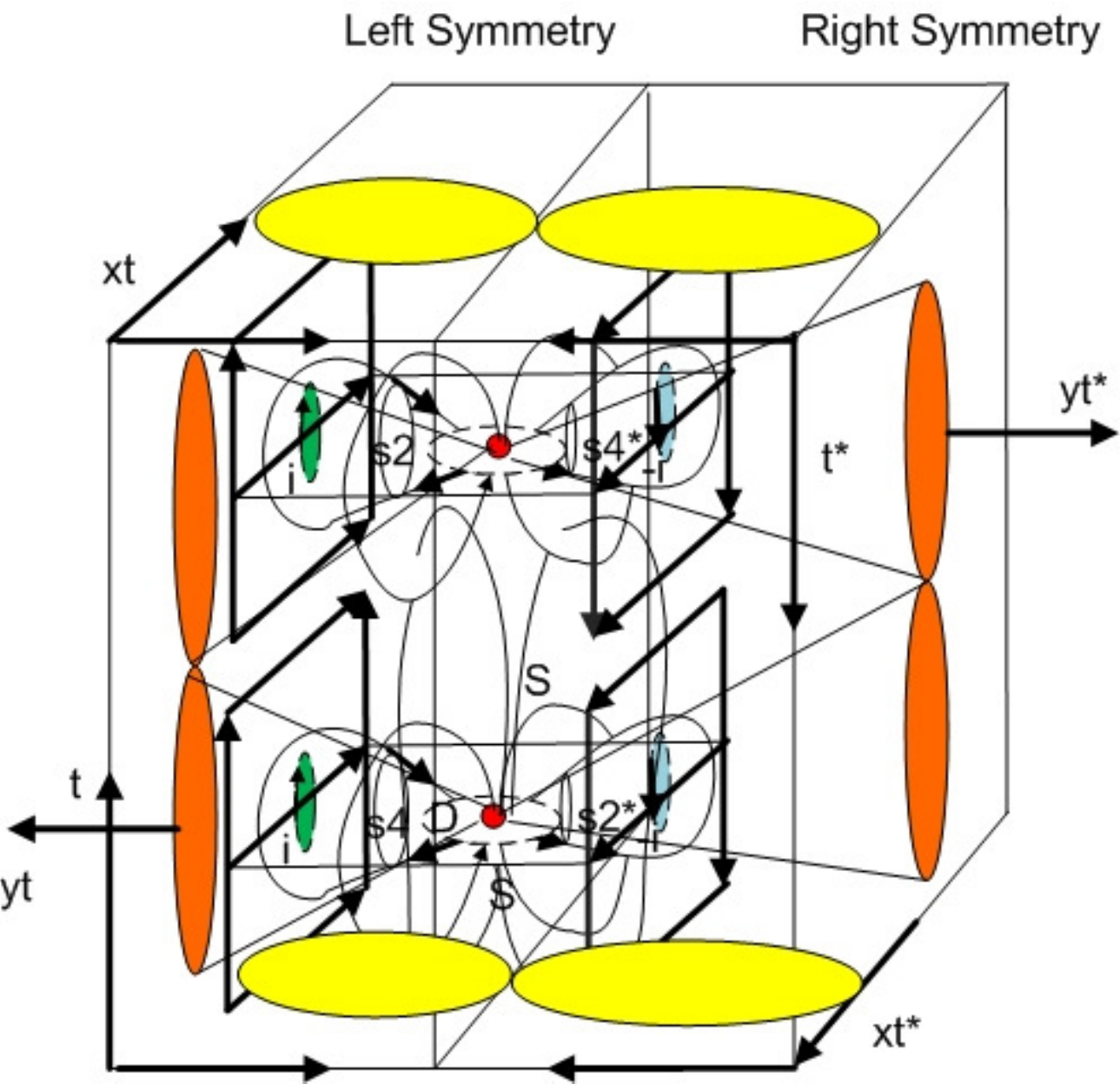,width=5cm}
\epsfig{file=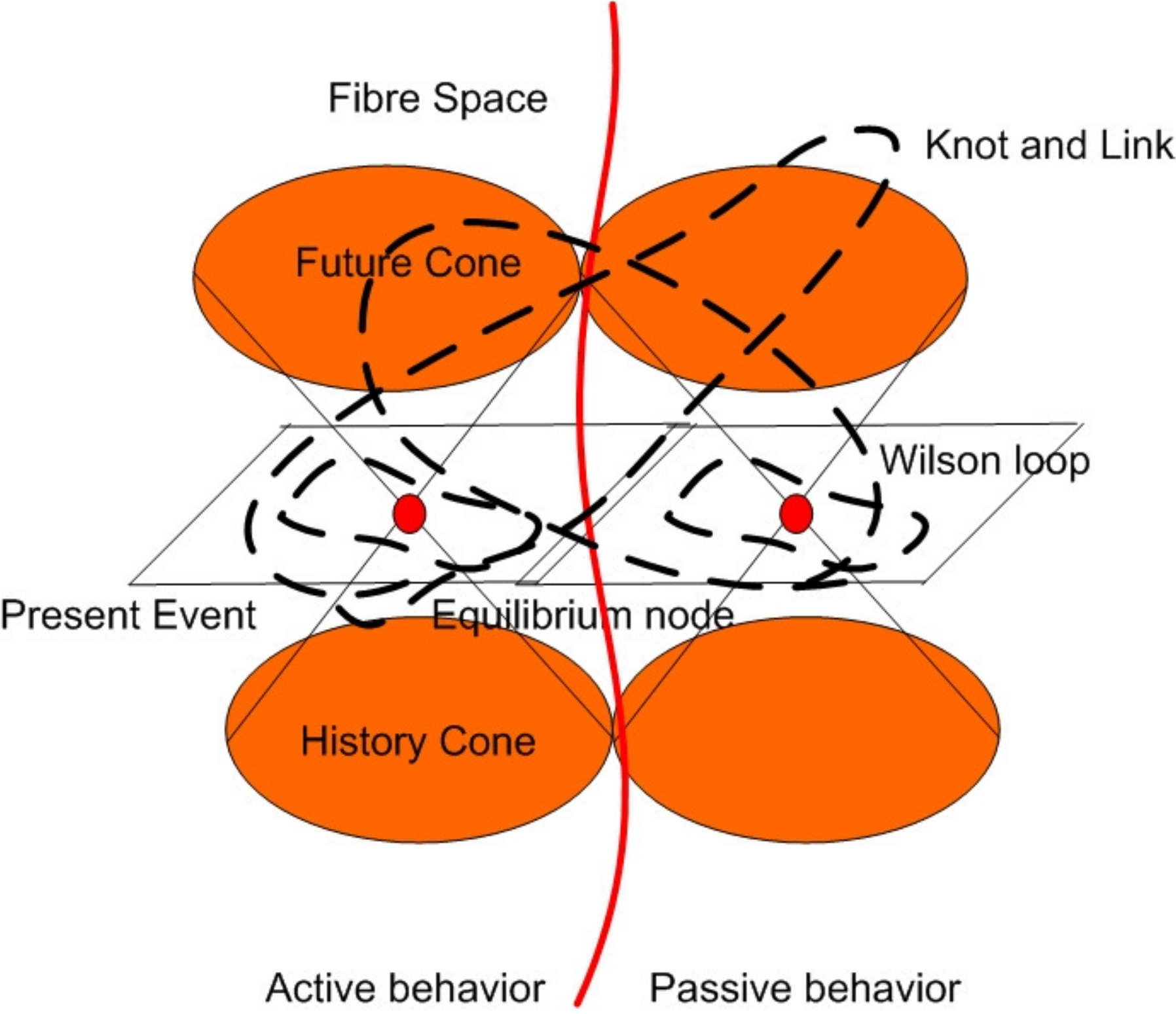,width=5cm}
 \caption{  Picture shown on the right is a  knot model of the time series data in the left and right link of the Wilson loop of the behavior of the trader in the supersymmetry with an extra dimension. The picture on the left is a  model of a supersymmetry theory $E_{8}\times E_{8}$ for the financial time series data.
\label{unify}}
\end{figure}

We use a Laurent series to decompose the risk field into the left excess demand, $D$ and  the right excess demand $D^{\ast}$ in the hidden space  knot the Laurent polynomial with the coefficient in the Wilson loop of the behavior of the trader over the knot of the market cocycle $(\alpha,\beta)$ ,$W_{\alpha,\beta}(A_{\mu})$ by
\[\sigma_{t}^{2}=\oint_{H^{2}(\mathcal{O}^{p+q}_{X})}   \frac{\Pi W_{\alpha,\beta} (A_{\mu})}{D-D_{0}} dD
+\oint_{H^{2}(\mathcal{O}^{p+q}_{X^{\ast}})}   \frac{\Pi W_{\alpha,\beta}(A_{\mu}) }{D^{\ast}-D_{0}^{\ast}} dD^{\ast}+ \oint_{H^{2}(\mathcal{O}^{p+q}_{Y})}   \frac{\Pi W_{\alpha,\beta} (A_{\mu})}{S-S_{0}} dS
+\oint_{H^{2}(\mathcal{O}^{p+q}_{Y^{\ast}})}   \frac{\Pi W_{\alpha,\beta}(A_{\mu}) }{S^{\ast}-S_{0}^{\ast}} dS^{\ast},   \]
where $(S,S^{\ast})$ is a pair of the supply and the hidden supply and $(D,D^{\ast})$ is a pair of the demand and a hidden demand field.
The main problem of GARCH(1,1) is that return is not in a log scale but we have a constant Euclidean distance. In this work, we change the metric to the hyperbolic distance in order to solve a volatility of the cluster phenomena. We found a complete solution in the complex time scale with a conjugate solution in the minus time scale. We interpret the results with the index cohesive force. We called the volatility in the inert hidden time scale the hidden fear field or the dark volatility.

 We  model the financial market in the unified theory of $E_{8}\times E_{8}^{\ast}$ (Figs. \ref{unify}) under the mathematical structure of the Kolmogorov space underlying the time series data. Let $M\in\mathbb{H}P^{1}/Spin(3)$ be a market with the supply and demand sides and $M^{\ast}:\rightarrow\mathbb{H}P^{1}/Spin(3)\rightarrow \mathbb{H}P^{1}$ be a dual market of the behavior of the noise trader, $\sigma$ and fundamentalist $f$. Then we prove an existence of the general equilibrium point in the market by using equivalent classes of the supply $[s]$ and the demand $[d]$ in the complex projective space $[s],[d]\in \mathbb{C}P^{1}$.
 The Kolmogorov space $S^{7}=\mathbb{H}P^{1}\simeq E_{8}$ can use explicit states of the financial market and its dual $\mathbb{H}P^{1\ast}/spin(3)\simeq E_{8}^{\ast}$ uses dark states in the financial market. It is a ground field of the first knot cohomology group tensor with the first de Rahm cohomology group
 $H_{1}(x_{t};[G,G^{\ast}];\mathbb{H}P^{1}/spin(3))\otimes H^{1}(M;\mathbb{H}P^{1\ast}/spin(3)).$ We induce free duality maps $\xi:H_{1}(x_{t};[G,G^{\ast}];\mathbb{H}P^{1}/Spin(3))\rightarrow H^{1}(M;\mathbb{H}P^{1\ast}/spin(3)) $ as the market states $E_{8}\times E_{8}^{\ast}$ in the grand unified theory model of the financial market.
For the simplicity, we let a piece of the surface cut out from the Riemann sphere with a constant curvature $g_{ij}$ defined as a parameter of a D-brane field basis with the same orientation as a definition of the supply and demand in the complex plane (flat Riemann surface). Now we blend the complex plane to the hyperbolic coordinate in 8-hidden dimensions of $\Psi=(\Psi_{1},\cdots \Psi_{8}^{\ast})\in S^{7}$, 8 states in the explicit form.

\subsection{Yang-Mills equation for Financial Market}
In this section, we explain the source of  de Rahm coholomogy of the financial market. The mathematical structure in this section is related to the Chern-Simons theory of so- called gravitational field in 3 forms of the connection $F^{\bigtriangledown}\in \Omega^{2}(T_{x}M\otimes T_{x}^{\ast}M)$ over 3 vector fields in the financial market: $\Psi_{i}\in M$ a market state field, $\mathcal{A}_{i}$ an agent behavior field and $[s_{i}]$ a field of physiology of financial time series data over Kolmogorov space. We simplify the Chern-Simons theory to a suitable form to easy understand this section and also can just simplify the prototype of the real and useful new cohomology theory for the financial market. Let $\mathcal{A}=(A_{1},A_{2},A_{3})$ be a market micro potential field of the agent or the behavior of the trader in the market. Let $\Psi_{i}^{(\mathcal{S},\mathcal{D})}([s_{i}],[s_{j}])$ be a field of the market $8-$ states of the equivalent class of the physiology of the time series $[s_{i}],[s_{j}]$.

In the financial market microstructure,  we define a new quantity in microeconomics induced from the interaction of the order submission from the 
 supply and demand side of the orderbook, so-called market microvector potential
 field or the connection in the Chern-Simons theory for the financial time series $ \mathcal{A}(x,t;g_{ij},\Psi_{i}^{(\vec{\mathcal{S}},\vec{\mathcal{D}})},s_{i}(x_{t}))$, where  $\Psi_{i}^{(\vec{\mathcal{S}},\vec{\mathcal{D}})}$ is a state function for the supply $\vec{\mathcal{S}}$ and the demand
$\vec{\mathcal{D}}$ of the financial market. Let $ \Psi_{i}^{(\vec{\mathcal{S}},\vec{\mathcal{D}})}(F_{i})$
 be a scalar field induce  from news of market factor $F_{i},i=1,2,3,4$. Let $x_{t}(\mathcal{A},x^{\ast},t^{\ast})$ be a financial time series induce from field of market behavior of trader. Where $x^{\ast}$ and $t^{\ast}$ are hidden dimension of time series data in Kolmogorov space.

In equilibrium state, the sequence is exact with $\partial^{2}=0.$ We have demand in equilibrium state $D=\partial^{2}A=0$ as property of short exact sequence in which can induce infinite cohomology sequence of sphere as sheave sequence of financial time series data.

\begin{Definition}
Let $S$ be a supply  potential field   be defined by rate of change of  supply side of market potential fields of behavior trader in induced field of behavior of  agent $\mathcal{A}$ in stock market.

\begin{equation}
\vec{\mathcal{S}}=-\frac{\partial \mathcal{A}}{\partial [s_{4}]}
\end{equation}

\begin{equation}
\vec{\mathcal{NS}}=-i\frac{\partial \mathcal{A}}{\partial [s_{4}]}
\end{equation}
\begin{equation}
\vec{\mathcal{D}}=\frac{\partial \mathcal{A}}{\partial [s_{2}]}
\end{equation}
\begin{equation}
\vec{\mathcal{ND}}=+i\frac{\partial \mathcal{A}}{\partial [s_{2}]}
\end{equation}
with
\begin{equation}
\mathcal{A}=(A_{1}(f),A_{2}(\sigma),A_{3}(\omega))
\end{equation}
\end{Definition}

\begin{Definition}
Let $\vec{\mathcal{D}}$ be a demand  potential field (analogy to magentic field). It is
 defined by rate of change of market potential fields in induce field of $\vec{\mathcal{S}}$,

\begin{equation}
\vec{\mathcal{D}}=\bigtriangledown \times \vec{\mathcal{S}}.
\end{equation}

 \end{Definition}

\begin{Definition}
Let $F_{\mu\nu}$ be a stress tensor for market microstructure
 with component $A_{\mu}$ of market potential field(analogy with connection one form in Chern-Simons theory) defined by the rank 2 antisymmetric tensor field,

\begin{equation}
F_{\mu\nu} =\partial_{\mu}A_{\nu}-\partial_{\nu}A_{\mu},
\end{equation}
\begin{equation}
\frac{d\mathcal{A}}{d[s_{2}]}=S_{\nu}=F_{\mu\nu}\Psi^{\nu}, \hspace{0.5cm}  \frac{d\mathcal{A}}{d[s_{4}]}=  D_{\mu}=\epsilon_{\mu\nu}^{\kappa\lambda}F_{\kappa\lambda}\Psi^{\nu}.
 \end{equation}

\end{Definition}

These 3 vector  fields of behavior trader play role of 3 form in tangent of complex manifold. It is an element of section $\Gamma$ of manifold of market,
$ \mathcal{A}\in \Gamma(\wedge^{3}T^{\ast}_{x}M\otimes \mathbb{H})= \Omega^{3}(M).$

Let consider 3-differential form of behavior of trader as market potential field $\mathcal{A}$,

\begin{equation}
d\mathcal{A}=\sum_{ijk=1,2,3}F_{ijk}dA_{i}\wedge dA_{j}\wedge dA_{k}
\end{equation}
\begin{equation}
\begin{CD}
\cdots @<<<\Omega^{3}(M) @<{\text{d}}<< \Omega^{2}(M) @<{\text{d}}<< \Omega^{1}(M)@<{\text{d}}<< \Omega^{0}(M)@<{\text{d}}<< 0 \\
    @.   @VVV @VVV @VVV @VVV @.\\
\cdots @<<<C_{3}(X_{t}) @>\partial>> C_{2}(X_{t}) @>\partial>> C_{1}(X_{t})@>\partial>> C_{0}(X_{t})@>\partial>> 0
\end{CD}
\end{equation}
with a market state $\Psi^{(S,D)}\in C^{1}(M).$
A de Rahm cohomology for financial market is defined by using equivalent class over these differential form. The first class so called market cocycle $Z^{n}(M)=\{f^{\ast}: C^{n}(M)\rightarrow C^{n+1}(M)  \}$ with commutative diagram between covariant and contravariant functor,

\begin{equation}
\begin{CD}
\cdots @<<<H^{3}(M) @<{\text{d}}<< H^{2}(M) @<{\text{d}}<< H^{1}(M) @<{\text{d}}<< H^{0}(M)@<{\text{d}}<<0\\
@.  @VV\xi V @VV\xi V @VV\xi V @VV\xi V @VV\xi V\\
\cdots @<<<H_{3}(X_{t}) @>\partial>> H_{2}(X_{t}) @>   \partial>> H_{1}(X_{t})@>   \partial>> H_{0}(X_{t})@>   \partial>> 0\\
@. @VV\nu V @VV\nu V @VV\nu V @VV\nu V @VV\nu V\\
\cdots @<<<H^{3}(x_{t}:[G,G^{\ast}]) @<{\text{d}}<< H^{2}(x_{t}:[G,G^{\ast}]) @<{\text{d}}<< H^{1}(x_{t}:[G,G^{\ast}]) @<{\text{d}}<< H^{0}(x_{t}:[G,G^{\ast}])@<{\text{d}}<<0\\
 \end{CD}
 \end{equation}

It is a kernel of co-differential map between supply and demand to market potential field.

\begin{equation}
Ker(d:\Omega^{2}(M)\rightarrow \Omega^{3}(M)), d:[s_{i}](S,D)\mapsto \mathcal{A}(A_{1},A_{2},A_{3}),d[s_{i}](S,D)=0=\mathcal{A}
\end{equation}
Where $[s_{i}](S,D)$ is an equivalent class of physiology of time series data. The boundary map of cochain of market $\Omega^{2}(M)$ is a second equivalent class used for modulo state,

\begin{equation}
Im(d:\Omega^{1}(M)\rightarrow \Omega^{2}(M), d:\Psi^{(S,D)}\mapsto [s_{i}](S,D).
\end{equation}

\begin{Definition}
The de Rahm cohomology for financial time series is an equivalent class of second cochain of market
\begin{equation}
H_{DR}^{2}(M)=Ker(d:\Omega^{2}(M)\rightarrow \Omega^{3}(M))/Im(d:\Omega^{1}(M)\rightarrow \Omega^{2}(M)).
\end{equation}
\end{Definition}
The meaning of new defined mathematical object is use for measure a market equilibrium in algebraic topology approach.

The section of manifold induce a connection of differential form. The connection typically gravitational field in physics. In finance, connection use for measure arbitrage opportunity.
The connection allow us to use the Peterson-Codazzi equations of Killing form of parallel  transport of geodesic curve as hidden equilibrium equation for financial market over Riemannan surface of market,

Let $A_{\nu}=F_{\nu}^{\bigtriangledown}\in \Omega^{2}(T_{x}M\otimes T_{x}^{\ast}M), \bigtriangledown:\Omega^{0}(M)\rightarrow \Omega^{1}(M)$.Let $\Psi_{i}\in M: \mathbb{H}P^{1}/Spin(3)\rightarrow S^{7},[s_{i}]\in Physio(x_{t}):=G:\mathbb{C}P^{1}=S^{3}/S^{1}\rightarrow S^{3}=\mathbb{H}.$ Where $G$
 is a Lie algebras of predictor and $G^{\ast}$ is a predictant. $\Psi_{i}^{\ast} $ is dual basis of
market state  $\Psi_{i}$. We have the Peterson-Codazzi equations of arbitrage opportunity as Jacobian flow $g_{ij}$ by

\begin{equation}
 g(\bigtriangledown_{\Psi}[s_{i}^{\ast}],[s_{i}])+g(\Psi,\bigtriangledown_{[s_{i}]}\Psi^{\ast})=0.
\end{equation}
where $\bigtriangledown$ is a connection of time series data over financial manifold.
We get an equation of equilibrium point of traders hold when $\lambda$ is eigenvalue of arbitrage $g$.
The covariant derivative allow  us to measure the rate of change of 3 fields in financial market.
The rate of change of arbitrage opportunity behavior field of trader $A_{i}$ with respect to rate of change of physiology of time series data $d[s_{i}]=[s_{i}^{\ast}]-[s_{i}]$, the error between predictor and predictant is given by
$\bigtriangledown_{\Psi}[s^{\ast}]=0$.
The meaning of term $\bigtriangledown_{\Psi}[s^{\ast}]$ in the Peterson-Codazzi equations of arbitrage opportunity is the change of market state with respect to expectation field of physiology $[s^{\ast}]$.

\subsection{ Grothendieck Cohomology  for Financial Time Series}
Let Minkowski space of light cone in time series data be compose of  2 cones embedded into Euclidean plane in pointed space of time series as equilibrium node.
 The upper cone is for superspace of supply $X_{t}$ as Kolmogorov space in time series data. The down cone is superspace of demand side of market, $Y_{t}$.
We assume that in unit cell of non-Euclidean plane of time series data  composed of 2 sheet with left and right hand supersymmetry. It is superposition  to each others in opposite direction with
left hand and right supersymmetry of hidden direction of time $dt$ and reversed direction of time  scale $dt^{\ast}$ in which observe cannot be notice from outside the system.
The quantization of hidden state of behavior of trader in superspace of time series data is an unoriented supermanifold with ghost field and anti-ghost field in side of Dbrane and anti-Dbrane sheet of normal distribution in Minkowski space in time series data. We define a new
lattice theory with ghost pairs state of 2 pairs of supply and demand ghost field $(\Psi_{L},\Psi_{R})$

\begin{Definition}
Let $\Psi_{R}(D_{i}^{\ast})$ be a market right  state of demand side over behavior of trader $[A_{\mu}]$.
Let $\Psi_{L}(S_{i}^{\ast})$ be a market right  state of supply side over behavior of trader $[A^{\mu}]=[A_{\mu}^{\ast}]$.
Let $[A_{\mu}]\in \mathcal{A}^{G}$ be a space of gauge field of connection of behavior of traders in financial market.
 let $[A_{\mu}^{\ast}]\in \mathcal{A}^{\ast}_{G}$ be a dual space of gauge field of
connection of a genetic code of host cell define by
\begin{equation}
\Phi(A_{\mu})^{+}_{\Psi_{R} }=\sum_{i=1}^{n}[g_{ji},A_{\mu}]|\Psi_{R}(D_{i}^{\ast})> =\sum_{i=1}^{n}{\sigma_{t}^{\ast}}^{2}|\Psi_{R}(x^{\ast})>
\end{equation}
and
 \begin{equation}
\Phi(A_{\mu})^{-}_{\Psi_{L}}=\sum_{i=1}^{n}[g^{ji},A^{\mu}]|\Psi_{L}(S_{i}^{\ast})> =\sum_{i=1}^{n}{\sigma_{t}^{\ast}}^{2}|\Psi_{L}(y^{\ast})>
\end{equation}

 if there exist a market transition state  shift as dark volatility $\sigma
_{t}^{\ast}$ with gauge group of evolution behavior of trader $G$ such that,
\begin{equation}
 [ \Psi_{R}(x^{\ast})^{+}, \Psi_{L}(y^{\ast})^{-}]=\frac{\partial p_{t}(S^{\ast},D^{\ast})
}{\partial [s_{t}^{\ast}(t)]}
\end{equation}

with  ${{\sigma}^{\ast}}^{2}_{t}\in \mathcal{A}^{G}\otimes \mathcal{A}^{\ast}_{G}$.
\end{Definition}

The coupling  of interaction of supernormal distribution induced from behavior of trader as  pairs of ghost field and anti-ghost field can be classified by using link operator  in knot theory, with link state $<L>_{0},<L>_{+}$ and  $<L_{-}>$. A unoriented superdistrubtion  is in knot state with left and right link operator associate with left and right supersymmetry in chiral ghost field of supply and demand of market superstate.
We work in level of Kolmogorov space. Let $X_{t}$ be Kolmogorov space of supply gauge field with
$S_{t}\in X_{t}$ and $Y_{t}$ be Kolmogorov space of demand gauge field with $D_{t}\in Y_{t}$.
Let a momentum space of orbital of quantum price structure in supermanifold be coordinated by $(k_{x},k_{y})\in X_{t}/Y_{t}|_{L}$,we can choose $X_{t}/Y_{t}:=\mathbb{H}P^{1}$, with twistor break a chiral supersymmetry from left to right in mirror space  by

\begin{equation}
 \mathcal{O}^{p,q}_{X_{t}/Y_{t}|_{L}} \stackrel{<L>_{+}}{\rightarrow} \mathcal{O}^{p,q}_{Y_{t}^{\ast}/X_{t}^{\ast}|_{R}}
\end{equation}
\begin{equation}
  \mathcal{O}^{p,q}_{Y_{t}^{\ast}/X_{t}^{\ast}|_{R}}\stackrel{<L>_{-}}{\rightarrow} \mathcal{O}^{p,q}_{ X_{t}/Y_{t}|_{L}}
\end{equation}

with skein relation over knot of Laurent polynomial $ q=e^{\frac{2\pi i}{n+k}}$,
\begin{equation}
-q^{\frac{n}{2}}<L>_{+}
+(q^{\frac{1}{2}}  - q^{-\frac{1}{2}} )<L>_{0}+q^{-\frac{n}{2}}<L>_{-}=0.
\end{equation}

The dual behavior trader pair is $(k_{x}^{\ast},k_{y}^{\ast})\in Y_{t}^{\ast}/X_{t}^{\ast}.$ We have a short exact sequence induced infinite sequence of Grothendieck cohomology for financial time series data

\begin{equation}
0\rightarrow \mathcal{O}^{p,q}_{\mathbb{Z}/2} \rightarrow  \mathcal{O}^{p,q}_{X_{t}}  \rightarrow  \mathcal{O}^{p,q}_{Y_{t}}\rightarrow  \mathcal{O}^{p,q}_{X_{t}/Y_{t}}\rightarrow  \mathcal{O}^{p,q}_{Y_{t}^{\ast}/X_{t}^{\ast}}\rightarrow  \mathcal{O}^{p,q}_{Y_{t}^{\ast}} \rightarrow  \mathcal{O}^{p,q}_{X_{t}^{\ast}}\rightarrow  \cdots     \nonumber
\end{equation}

\begin{equation}
\rightarrow H^{1}(\mathcal{O}^{p,q}_{\mathbb{Z}/2 }) \rightarrow H^{1}(\mathcal{O}^{p,q}_{X_{t}})  \rightarrow H^{1}(\mathcal{O}^{p,q}_{Y_{t}})\rightarrow
H^{1}(\mathcal{O}^{p,q}_{X_{t}/Y_{t}})\rightarrow H^{1}(\mathcal{O}^{p,q}_{Y_{t}^{\ast}/X_{t}^{\ast}})\rightarrow H^{1}(\mathcal{O}^{p,q}_{Y_{t}^{\ast}}) \rightarrow H^{1}(\mathcal{O}^{p,q}_{X_{t}^{\ast}})\rightarrow       \nonumber
\end{equation}

\begin{equation}
\rightarrow H^{2}(\mathcal{O}^{p,q}_{\mathbb{Z}/2 }) \rightarrow H^{2}(\mathcal{O}^{p,q}_{X_{t}})  \rightarrow H^{2}(\mathcal{O}^{p,q}_{Y_{t}})\rightarrow
H^{2}(\mathcal{O}^{p,q}_{X_{t}/Y_{t}})\rightarrow H^{2}(\mathcal{O}^{p,q}_{Y_{t}^{\ast}/X_{t}^{\ast}})\rightarrow H^{2}(\mathcal{O}^{p,q}_{Y_{t}^{\ast}}) \rightarrow H^{2}(\mathcal{O}^{p,q}_{X_{t}^{\ast}})\rightarrow  \cdots   \nonumber
\end{equation}

\begin{equation}
\rightarrow H^{n}(\mathcal{O}^{p,q}_{\mathbb{Z}/2 }) \rightarrow H^{n}(\mathcal{O}^{p,q}_{X_{t}})  \rightarrow H^{n}(\mathcal{O}^{p,q}_{Y_{t}})\rightarrow
H^{n}(\mathcal{O}^{p,q}_{X_{t}/Y_{t}})\rightarrow H^{n}(\mathcal{O}^{p,q}_{Y_{t}^{\ast}/X_{t}^{\ast}})\rightarrow H^{2}(\mathcal{O}^{p,q}_{Y_{t}^{\ast}}) \rightarrow H^{n}(\mathcal{O}^{p,q}_{X_{t}^{\ast}})\rightarrow  \cdots
\end{equation}

Let  a behavior of traders be a ghost pairs $\Phi^{\pm}:= (\Psi^{R},\Psi^{L})$ with left supersymmetry be

\begin{equation}
\Psi_{L}:= \Psi_{L}^{ 0\rightarrow g_{1}^{L}\stackrel{d_{1}}{\rightarrow}  g_{2}^{L}\stackrel{d_{2}}{\rightarrow} g_{3}^{L}\rightarrow \cdots \stackrel{d_{n}}{\rightarrow}g_{n}^{L}\rightarrow 0}
\end{equation}
where $g_{i}^{L}$ is an adjoint map of market chain of cocycles.
\begin{equation}
g_{i}^{L}:
\mathcal{O}^{p,q}\{0 \rightarrow  \mathbb{Z}/2\rightarrow X_{t}  \rightarrow Y_{t}\rightarrow X_{t}/Y_{t} \rightarrow \cdots\} \stackrel{<L>_{-}}{\rightarrow}
\mathcal{O}^{p,q}\{0\rightarrow  Y_{t}^{\ast}/X_{t}^{\ast}\rightarrow Y_{t}^{\ast} \rightarrow X_{t}^{\ast}\rightarrow  \cdots \}
\end{equation}

  Let a supply side of market  as a right supersymmetry be
\begin{equation}
\Psi_{R}:= \Psi_{R}^{ {g_{1}^{\ast}}^{R}\leftarrow  {g_{2}^{\ast}}^{R}\leftarrow {g_{3}^{\ast}}^{R}\leftarrow
\cdots {g_{n}^{\ast}}^{R}}
\end{equation}
\begin{equation}
{g_{i}^{\ast}}^{R}:
\mathcal{O}^{p,q}\{\cdots\leftarrow X_{t}  \leftarrow Y_{t}\leftarrow X_{t}/Y_{t} \leftarrow 0\} \stackrel{<L>_{+}}{\leftarrow}
\mathcal{O}^{p,q}\{\cdots\leftarrow  Y_{t}^{\ast}/X_{t}^{\ast}\leftarrow Y_{t}^{\ast} \leftarrow X_{t}^{\ast}\leftarrow  \mathbb{Z}/2  \leftarrow 0 \}
\end{equation}
with
\begin{equation}
Ad_{\Psi^{L}}{\Psi^{\ast}}^{R}=  \{\Psi^{L},{\Psi^{\ast}}^{R}\} =0
\end{equation}
The inversion property of CPT  produce parity inversion as ghost field and anti-ghost field pairs.
Let ghost field pair be $\Phi^{+}_{\Psi^{\ast}_{R}}(A_{\mu})$ with arbitrage opportunity as a connection $A_{\mu}$ and anti-ghost field  pair be $\Phi^{-}_{\Psi_{L}}(A_{\mu})$ with partition map
\begin{equation}
p(\Phi^{-}_{\Psi_{L}(x_{t}^{\ast})}(A_{\mu}))+ p(\Phi^{+}_{\Psi^{\ast}_{R}(y_{t})}(A_{\mu}))=-1
\end{equation}
where the parity map separate hidden supersymmetry of left right in supply demand equilibrium  node represent as market general equilibrium point between supply and demand gauge field in 2 cones of with price surfaces by $p: \{ \Phi^{-}_{\Psi_{L}}(A_{\mu}),\Phi^{+}_{\Psi_{R}}(A_{\mu})\}\rightarrow \mathbb{Z}_{2}=\{-1,1\}$ with $p( \Phi^{-}_{\Psi_{L}(x_{t}^{\ast})}(A_{\mu})))=1-\Phi^{+}_{\Psi_{L}(x_{t})}(A_{\mu}))$ and $p( \Phi^{-}_{\Psi_{R}(y_{t}^{\ast})}(A_{\mu})))=1-\Phi^{+}_{\Psi_{L}(y_{t})}(A_{\mu}))$ as time reversal partity operator in moduli state space model of financial market.

Let  define quantum price with for time reversal wave function by $\Psi(x^{\ast})= \Psi(x_{t})^{t}=e^{-ik_{x_{t}}x}$. We have

\begin{figure}[!t]
\centering
\epsfig{file=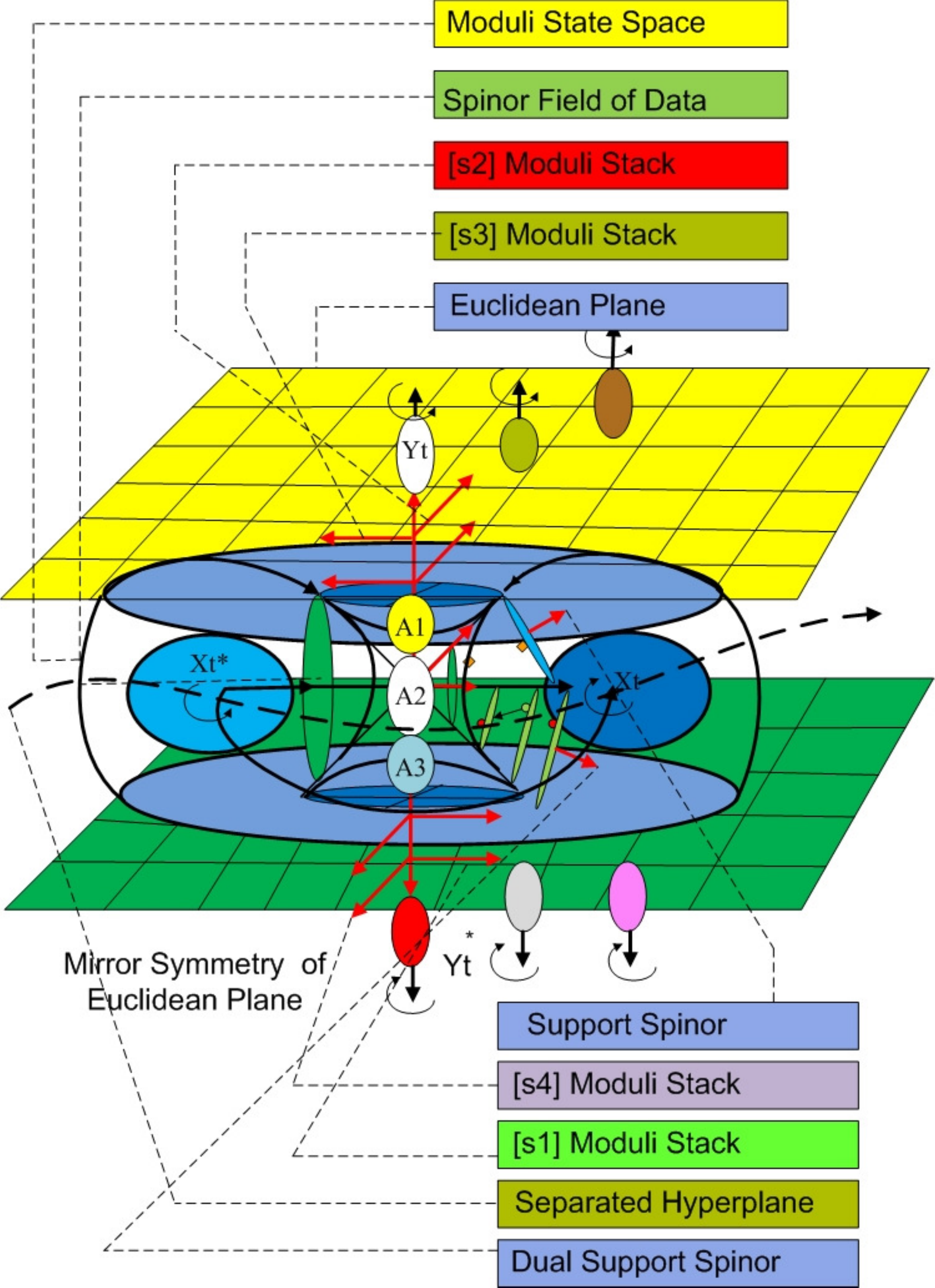,width=8cm}
 \caption{  Picture shown superspace in time series data with moduli stack $[s_{i}].$ The cone inside superspace is a Minkowski space with light cone for time series data.  We have spinor field in time series data as support spinor machine in this model.
\label{ssm}}
\end{figure}

 The superspace in time series data use to measured (see fig. \ref{ssm} for detail) the dual coupling behavior field derived from Laurent series over 4 hidden momentum spaces, $X_{t},Y_{t},X_{t}^{\ast},Y_{t}^{\ast}$ with both positive and negative coefficient of knot polynomial represent optimistic and pessimistic fundamentalist and noise traders behavior field. The knot behavior of  behavior induced an arbitrage opportunity as a  connection over Wilson loop represent market phase transition with $(\alpha ,\beta )$ market  cocycle in $W_{\beta,\alpha}(A_{\mu})$.

 Let a momentum coordinate of price quantum  be in functional coordinate over path integral of Wilson loop
along loop space of ghost field behavior  pairs. We have
\begin{equation}
k_{x}=\frac{i}{x}\int_{0}^{\Psi_{R}}\frac{\Pi W_{\alpha,\beta}(A_{\mu})}{x^{\ast}-x^{\ast}_{0}}dx^{\ast},
\end{equation}
\begin{equation}
k_{y}=\frac{i}{y}\int_{0}^{\Psi_{L}}\frac{\Pi W_{\alpha,\beta}(A_{\mu})}{y^{\ast}-y^{\ast}_{0}}dy^{\ast}
\end{equation}
\begin{equation}
k_{x^{\ast}}=\frac{i}{x^{\ast}}\int_{0}^{\Psi_{R}}\frac{\Pi W_{\alpha,\beta}(A_{\mu})}{x-x_{0}}dx,
\end{equation}
\begin{equation}
k_{y^{\ast}}=\frac{i}{y^{\ast}}\int_{0}^{\Psi_{L}}\frac{\Pi W_{\alpha,\beta}(A_{\mu})}{y-y_{0}}dy
\end{equation}
A left pairs of behavior field represent supply side is

\begin{equation}
\Phi^{+}_{\Psi_{L}  }=c_{k_{x}}e^{2\pi i k_{x}}-c_{k_{y}}e^{2\pi i k_{y}}
\end{equation}
and right pair represent demand side of financial market with
\begin{equation}
\Phi^{-}_{\Psi_{R}  }=c_{k_{x^{\ast}}}e^{-2\pi i k_{x^{\ast}}}+c_{k_{y^{\ast}}}e^{-2\pi i k_{y^{\ast}}}
\end{equation}
The equilibrium node $\{(\ast_{x},\ast_{y})\}$in Minkowski market light cone
$Cone(x^{\ast},y^{\ast};pp^{\ast}(k^{\ast})):=\{ {x^{\ast}}^{2}-{y^{\ast}}^{2}=||pp^{\ast}(k^{\ast})||^{2}=1\}$ is defined by cone
\[\{(\ast_{x},\ast_{y})\} \in Cone(x,y;pp(k))\]
and where $pp^{\ast}(k^{\ast})$ is a superdistribution of price quantization and $x^{\ast},y^{\ast}$ are a hidden demand and hidden supply $,D^{\ast},S^{\ast}$.
We define Galois connection to map between 2 dual Minkowski cones as market equilibrium path by $ \{(\ast_{x^{\ast}},\ast_{y^{\ast}})\}\in Cone(x^{\ast},y^{\ast};pp^{\ast}(k^{\ast}))$ with
\begin{equation}
Cone(x,y;pp(k))\stackrel{f}{\rightarrow} Cone(x^{\ast},y^{\ast};pp^{\ast}(k^{\ast})),Cone(x^{\ast},y^{\ast};pp^{\ast}(k^{\ast}))\stackrel{g}{\rightarrow} Cone(x,y;pp(k))
\end{equation}
 \begin{equation}
\{(\ast_{x},\ast_{y})\} =g(f( \{(\ast_{x},\ast_{y})\} ),\{(\ast_{x^{\ast}},\ast_{y^{\ast}})\} =f(g( \{(\ast_{x}^{\ast},\ast_{y}^{\ast})\} )
\end{equation}

In supernormal distribution, we have take into account spin   as coupling field of behavior of trader appear as Wilson loop of Pauli matrix $f_{+}:=A_{\mu=1}:=-W_{\alpha,\beta}(\sigma_{z})$
as behavior of fundamentalist. The behavior of chatlist  is  defined by Pauli matrix
 $\sigma_{+}:=A_{\mu=2}:=\sigma_{y}$. The behavior of bias traders is an adjoint representation of fundamentalist and inversed of Wilson loop of chatlist
 $A_{3}:=\frac{1}{2i}[f,-W^{-1}(\sigma)].$ The optimistic behavior is denoted by $A_{\mu}$ and pessimistic is denoted by $A^{\mu}=(A_{\mu})^{\ast}$.
Hence
\begin{equation}
pp^{\ast}(k^{\ast}):=A_{\mu}\Phi^{+}_{{\Phi}_{R}}+A^{\mu}\Phi^{-}_{{\Phi}_{L}}
\end{equation}
with $||pp^{\ast}(k^{\ast})||^{2}=1$ and $||A_{\mu}\Phi^{+}_{{\Phi}_{R}}||^{2}=1, ||A^{\mu}\Phi^{-}_{{\Phi}_{L}}||^{2}=1.$
If $A_{\mu}$ is a Pauli matrix represented spinor field of behavior of trader. The spinor invariant property of
price quantum $\varphi_{t}$ in these normalization formula is a source of stationary process in superspace of time series data extended volatility to dark volatility by $||\varphi_{i}||^{2}=1$ to $||\sigma_{i}\varphi_{i}||^{2}=1$.

Let $S^{2}$ be financial market as a Riemann sphere of price momentum space of price quantum without singularity, glue up from dual side of supply and demand in market with   2  disjoint cones with 2 equilibrium  nodes. The market equilibrium node is composed of pairs opposite spinor field of behavior traders, $(\ast_{x},\ast_{y})$ as a  pointed space embedded inside Riemann sphere of energy surface $S^{2}$.

\begin{equation}
S^{2}=X_{t}\coprod X_{t}^{\ast}\coprod Y_{t}\coprod Y_{t}^{\ast}\coprod \ast_{x_{t}}\coprod \ast_{y_{t}}
\end{equation}

where $x_{t}\in X_{t},y_{t}\in Y_{t},x_{t}^{\ast}\in X_{t}^{\ast},y_{t}^{\ast}\in Y_{t}$ are Kolmogorov space for time series data.

This is a new kind of super mathematics theory in probability theory with new integral sign over tangent of supermanifold under Berezin coordinate transformation.
The structure of supergeometry of superpoint induce a superstatistic of hidden ghost field in financial market. We let $y_{t}$ be an observe variable in state space model and $x_{t}$ be hidden variable of state. We denote $\Phi_{i}(y_{t})\in \mathcal{A}$ be a ghost field and $\Phi_{i}^{+}(x_{t})$ be an anti-ghost field in finance with relation of its ghost number ,

\begin{equation}
gh(\Phi_{i}(y_{t}))+gh(\Phi_{i}^{+}(x_{t}))
=-1.
\end{equation}
We have a parity $p(\Phi_{i}(y_{t}))=1-p(\Phi_{i}^{+}(y_{t})) $ and $p(\Phi_{i}(x_{t}))=1-p(\Phi_{i}^{+}(x_{t})) $
 for both state and space variable in ghost field and anti-ghost field.

\subsection{ Volatility Clustering Phenomena and General Equilibrium in Financial Market}

 We separate the system of financial market into 2 parts, first is state part $X_{t}$ of hidden demand state  and second is a space part $Y_{t}$ of observation of supply  space of state space model. In this work, we use algebraic equation from algebraic topology and differential geometry as a main tool for define a new mathematical object for arbitrage opportunity in DSGE system of macroeconomics.

\begin{Theorem}
When market is in equilibrium , we have
\begin{equation}
s^{2}=0 \leftrightarrow H^{-14}(\mathcal{A},s)=0
\end{equation}
\end{Theorem}
{\em proof:} see  \cite{anomaly} .

We divide market into 2 separated sheet of Dbrane and
anti-Dbrane of embedded indifference curve of supply curve and  utility curve of demand.
 The interaction of 2 Dbrane induce from trade off between supply and demand as general equilibrium point. We define the Dbrane sheet of market is in real dimension and the anti self duality (AdS) of Dbrane to anti-Dbrane is induce from duality map from supply to demand.

 Let ordinary least square (OLS) in superspace in time series data be written by
\begin{equation}
  y_{t} =\alpha_{t} +\beta_{t} x_{t} +  \epsilon_{t}
\end{equation}

Take a    ghost  functor $\Phi_{i},\Phi_{i}^{+}:  X\rightarrow (\mathcal{A},s)\simeq [X,S^{\pm k}]$

\begin{equation}
  \Phi_{i}( (y_{t} - \alpha_{t})- \beta_{t} x_{t} \simeq  \epsilon_{t})
\end{equation}

 Let $\epsilon_{t}$ be real present shock from economics and $\epsilon_{t}^{\ast}$ be expected shock.  in the futures. In equilibrium, we assume steady stead of macroeconomics with no shock i.e. $\epsilon_{t}^{2}=0$. Let 0 be space of equilibrium and equivalent with moduli state space model of supply space $Y_{t}$ and demand space $X_{t}$ of market with $\epsilon_{t}^{2}=<\epsilon_{t},\epsilon_{t}^{\ast}>\simeq Y_{t}/X_{t}$. Consider short  exact sequence of macroeconomics in general equilibrium
with market risk $\beta_{t}$ of sudden shock in demand side and transfer into supply side of economics and let systematics risk $\alpha_{t}$ be a  shock in both side of with price sticky market.

   The moduli group $\mathbb{Z}_{2}$ define a state up and down of underlying financial time series data.
When market is in  equilibrium the short exact sequence will induce infinite exact sequence of market cocycle $\beta_{t}$ and $\alpha_{t}$,

In order to proof  the existence of dark volatility in financial market. We use tensor correlation field. Let $\tau$ be time lag of autocorrelation function
     of volatlity cluster,
\begin{equation}
C_{|\tau|} = Corr(|r_{t} |, |r_{t+\tau} |) =\frac{C}{\tau^{\beta}}
\end{equation}
with power law parameter $\beta\leq  0.5$. We define time lag of
power law function by, $ x^{\ast}(\tau, \beta) := \tau^{\beta}$.

Let $ds^{2} = x^{2} - {x^{\ast}}^{2}$
be a Minkowski metric
for GARCH(1,1). Let  $x$ be autocorrelation function $x =:C_{| r|}(\tau)$
of return with volatility cluster phenomena.
 Then then
there exist a dark volatility ${\sigma^{\ast}}_{t}^{2} < 0$ with minus sign in complex
 hidden time scale as a  homogeneous coordinate in quaternionic projective space $\mathbb{H}P^{1}$
 according to equation
 of projective hyperplane in Minkowski space $x^{2} - {x^{\ast}}^{2} = 1$.

 Let $\kappa$ be excess kurtosis, with $\sigma^{4} = \frac{\mu^{4}}{\kappa}-3.$
It is clear that $  \sigma^{2}=\pm \sqrt{  \frac{\mu^{4} }{\kappa}-3}$. Let us consider $x := C_{|r|}(\tau) =
Corr(|r_{t} |, |r_{t+\tau} |) = \frac{C}{\tau^{\beta}} =\frac{C}{x^{\ast}}$. So we have a rotated
hyperbola graph, $x x^{\ast} = C$ in with we can transform
by using rotational eigenvalue to normal hyperbola graph
$x^2-{x^{\ast}}^2 =  C$. If we normalize $C'$ to one.
If we use Euclidean distance in superspace in time series data with curvature
constant over Minkowski cone so we have no
arbitrage opportunity with time series is in stationary state.
Although, in parallel transport of  the connection
$A_{\mu}(t)$  is vary with time so we still have arbitrage opportunity in form of Wilson loop over market cocycle $(\alpha,\beta)$,$W_{\alpha,\beta}(A_{\mu})$.

In superspace of time series data $X_{t}/Y_{t}$, where $x:= C_{|r_{t}|}(\tau)\in X_{t}$
 and  $y:= C_{|r_{t+\tau}|}(\tau)\in Y_{t}$.We have tensor correlation as network of behavior traders, $x\otimes x^{\ast}, y\otimes y^{\ast}, x\otimes y, x^{\ast}\otimes y^{\ast} $ by

 \begin{equation}
 x\otimes y= C_{|r_{t}|}(\tau)\otimes C_{|r_{t+\tau}|}(\tau)\simeq Corr(|r_{t} |, |r_{t+\tau} |) = \frac{C}{\tau^{\beta}} = \frac{C}{x^{\ast}}.
 \end{equation}
We have a triplet of tensor correlation with memory in constant value $C$,

\begin{equation}
x^{\ast}(x\otimes y)=C.
\end{equation}
This triplet has properties analogous with the gravitational field in the triplet of the Lie algebras in the Nahm equation for the monopole or the instanton in the theoretical physics. We defined a new form of the Yang-Mills equation in the financial market as a Nahm equation in the financial market version as a main consequence of the proved of volatility clustering phenomena.
Let as denote $x:=D$, a demand field, $y:=S$ a supply field.
Let price be a moduli state space between the supply and demand $p_{t}=D_{t}/S_{t}.$

 Let $[D_{t}]:=T_{1}^{\mathcal{O}_{X}},[S_{t}]:=T_{2}^{\mathcal{O}_{Y}},[p_{t}]:=T_{3}^{\mathcal{O}_{X/Y}}$.
The Nahm equations for transition states in financial market are

\begin{equation}
\frac{dS_{t}}{d[s_{i}]}=[D_{t},p_{t}]
\end{equation}
\begin{equation}
\frac{dD_{t}}{d[s_{i}]}=[p_{t},S_{t}]
\end{equation}
\begin{equation}
\frac{dp_{t}}{d[s_{i}]}=[D_{t},S_{t}].
\end{equation}

 The three equations can be written  together by
\begin{equation}
\frac{dT_{i}^{\mathcal{O}_{(X,Y,X/Y)}}}{d[s_{i}]}=\frac {1}{2}\sum_{j,k}\epsilon_{ijk}[T_{j}^{\mathcal{O}_{(X,Y,X/Y)}},T_{k}^{\mathcal{O}_{(X,Y,X/Y)}}]=
\sum_{j,k}\epsilon_{ijk}T_{j}^{\mathcal{O}_{(X,Y,X/Y)}}T_{k}^{\mathcal{O}_{(X,Y,X/Y)}}.
\end{equation}
 With the modified Nahm equation for the interaction of the behavior of the traders from the supply and the demand sides in the financial market, we let  a pair of the co-states in the market from the pair of the supply and demand sides be modified Lax pair equations for the pairs of the market state in the financial market as the co-states between the fundamentalist and the chatlist with the optimistic and the  pessimistic forward looking of the price under the risk fear field ${\sigma_{t}^{\ast}}^{2}$.
It is trivial  that the  modified Nahm equations for biology can
 be written in the Lax pairs of the demand pairs, $D-$pairs  wave function and the supply pairs, $S-$ pairs wave function  as following.
Let
\begin{equation}
\Psi^{D-pairs}=T_{1}^{\mathcal{O}_{Y}}+iT_{2}^{\mathcal{O}_{X}},\Psi^{(D,S)-pairs}=-2iT_{3}^{\mathcal{O}_{X/Y}}, \Psi^{S-pairs}=T_{1}^{\mathcal{O}_{Y}}-iT_{2}^{\mathcal{O}_{X}}
\end{equation}
and
\begin{equation}
\Psi_{k,\mathcal{O}_{X}}^{\sigma_{t}^{2},D}=\Psi^{D-pairs}+k\Psi^{(D,S)-pairs} +k^{2}\Psi^{S-pairs},
\end{equation}
\begin{equation}
 \Psi_{k,\mathcal{O}_{Y}}^{{\sigma_{t}^{\ast}}^{2},S}=\frac{1}{2}\frac{d\Psi_{k,\mathcal{O}_{X}}^{\sigma_{t}^{2},D}}{d[s_{i}]}={\frac {1}{2}}\Psi^{(D,S)-pairs}+k \Psi^{S-pairs}
\end{equation}
where $k$ is a D-brane of the market coupling constant in the Chern-Simons current as a transition state in the quantum price
orbital, $J^{\mu=k}$.
Notice that dark volatility induces from the supply a short term ${\sigma_{t}^{\ast}}^{2}$ in market the transition state in the supply side of the financial market.
 The system of the Nahm equations is equivalent to the Lax pair equation for the financial market. The
  co-state $\Psi^{(D,S)-pairs}:=(\Psi_{k,\mathcal{O}_{X}}^{\sigma_{t}^{2},D}, \Psi_{k,\mathcal{O}_{Y}}^{{\sigma_{t}^{\ast}}^{2},S})$.
is a  transition between the co-state of the interaction of the behavior field of the trader from the supply  and the demand side in the financial market by the supply-demand  pairs in the general equilibrium of the market co-state  with $(\sigma_{t}^{2},{\sigma_{t}^{\ast}}^{2}) $ in the financial market.

 These system equations have a property of the Chern-Simons current in the time series data as the eigenvalues of the Dirac operator, $D$ for the financial market,
 \begin{equation}
 D\Psi^{(D,S)-pairs}=J^{\mu}\Psi^{(D,S)-pairs}
 \end{equation}

where the critical current with some threshold of the risk aversion for an arbitrage opportunity of the behavior of the trader is denoted by
${J^{\mu}}_{c}$. We have
 \begin{equation}
 \sin\Psi^{(D,S)-pairs}=\frac{J^{\mu}}{ {J^{\mu}}_{c}}=\Psi^{(D,S)-pairs}_{xx}-\Psi^{(D,S)-pairs}_{yy} = (\frac{\partial^{2}   }{\partial x^{2}}- \frac{\partial^{2}   }{\partial y^{2}})\Psi^{(D,S)-pairs}=D\Psi^{(D,S)-pairs}.
 \end{equation}
 We define a price as a transition state of an orbit state in the fibre space with the co-states of the transformation of the behavior field of the traders from the demand side of the market to the supply side by new  coordinates $(p_{t}^{\mathcal{O}_D}, p_{t}^{\mathcal{O}_S}) \in (\mathcal{O}_{X},\mathcal{O}_{Y})$ with
\begin{equation}
 p_{t}^{\mathcal{O}_D}=\frac{D_{t}+S_{t}}{2},\quad p^{\mathcal{O}_S}_{t}=\frac{D_{t}-S_{t}}{2},  {p_{t}^{\mathcal{O}_D}}^{\ast}=\frac{D_{t}^{\ast}+S_{t}^{\ast}}{2},\quad  {p_{t}^{\mathcal{O}_S}}^{\ast}=\frac{D_{t}^{\ast}-S_{t}^{\ast}}{2}.
\end{equation}
The equation can be transformed to
\begin{equation}
\Psi_{p^{\mathcal{O}_D}p^{\mathcal{O}_S}}^{D-pairs}=\sin \Psi^{D-pairs} ,
{\Psi^{\ast}_{{p^{\mathcal{O}_D}}^\ast{p^{\mathcal{O}_S}}^{\ast}}}^{S-pairs}=\sin {\Psi^{\ast}}^{S-pairs}.
\end{equation}
 The super stationary state of the demand pair solutions for the pairing of $(\mathcal{O}_D,\mathcal{O}_D^{\ast})$ and
$(\mathcal{O}_S,\mathcal{O}_S^{\ast})$ states in $(\mathcal{O}_S^{\ast},\mathcal{O}_D^{\ast})$ pairs  can be solved by
using the  Paterson-Godazzi equation for an arbitrage opportunity.
One part of the Lax pair equation  for an arbitrage is a Paterson Godazzi equation for the demand side which has the  pairing of $(\mathcal{O}_D,\mathcal{O}_D^{\ast})$ solution by
\begin{equation}
  \Psi^{D-pair}(D,S):=4\arctan
 e^{k\alpha (D-pS)+\sigma_{t} }, \Psi^{S-pair}(D^{\ast},S^{\ast}):=4\arctan
 e^{k\beta (D^{\ast}-p^{\ast}S^{\ast})+\sigma_{t}^{\ast} },
 \end{equation}
where $p_{t}=\frac{dD}{dt}/\frac{dS}{dt}=\frac{dD}{dS}$ is a parameter for the price as a transition market state by changing the coordinate of the $D$ coordination in the demand side of the superspace of the financial market and $S$ is the coordinate in the supply side. The new parameter ${\sigma_{t}^{\ast}}^{2}$ is a hidden risk fear field or the dark volatility in this work. It comes from  an evolutional factor of an adaptive behavior of the traders of the risk aversion and want to pursue an arbitrage opportunity.
We let the transition of the states between the active price and the passive price while price are matching in double auction market defined by the phase parameters of the market cocycle $(\alpha,\beta)$
$ \alpha^{2}=\frac{1}{1-p^{2}}, \beta^{2}=\frac{1}{1-{p^{\ast}}^{2}},$
where an extra parameter $k$ comes from the coupling constant of the behavior of the traders with the Chern-Simons current $J^{\mu}=J^{\mu}_{c}\sin\Psi $,

 The $(\mathcal{O}_D,\mathcal{O}_D^{\ast})$ pairs and $(\mathcal{O}_S,\mathcal{O}_S^{\ast})$-pairs  of the solution have a  positive root and a negative root for $ (\alpha,\beta)$ cocycle as  a twist in the variable to produce an arbitrage opportunity from the general equilibrium in the mean reversion as the super stationary state with the Chern-Simons current for the financial market. The phase shift  of the price quantum in the lax pairs
$ (\Psi^{D-pair},\Psi^{S-pair})$  takes the solution
$\Psi^{D-pair} =0=\Psi^{S-pair}$ to an adjacent with $ \Psi^{D-pair}=2\pi=\Psi^{S-pair}$ as a boundary condition for the price quantum with $    (\Psi^{D-pair},\Psi^{S-pair})=(0,0)(\textrm{mod}2\pi).$

\section{Result and Conclusion}

In this work we introduced  a new approach which was used to obtain the 
description of the dynamics of the market panic in the financial
world by using the  cohomology theory and the Yang-Mills theory over the traditional time series model, so-called GARCH(1,1).
Because the financial events understudied by the social sciences evidence cannot be
directly transported into the theory of Yang-Mills field for the financial market,
we need to add some more assumptions also to soften some of the
the assumptions of the Yang-Mills equations for the suitable use in the financial
market with the equilibrium point in the Minkowski spacetime.
We explained stylized facts in the financial time series data by using Yang-Mills equation for the financial market.
 We used Jacobian flow to construct the Minkowski metric with the hidden state in the time series data as the spinor field. We used ghost and anti-ghosts field in the Grothendieck cohomology to explain the interaction of the behavior of the traders from the supply and demand side of the market in an infinite cohomology sequence. We proved the power law in the volatility clustering phenomena by using the inversion invariant in the Yang-Mills theory. The result of this work is a new type of GARCH(1,1) model in the superspace of the time series data, so-called Minkowski GARCH(1,1). This theory is used to explain the spectrum of the nonstationary and nonlinear financial time series data as an analogy with the spectrum sequence of the behavior of the trader as pairs of the quantum field states in the underlying time series data in the spacetime geometry of the Kolmogorov space in time series data.
We proved all
the possibilities of the existence of the prior problem and all the possibilities
of the theorem to solve with the precise proving procedure. A collection of the theories for solving this problem we called the evolution feedback path.
The result of this solving can be
used to predict the financial time series with a very high precision
without end effect by using the mixture of filtering technique
and tool of theoretical physics.
In this paper, we discussed a stylized fact on the long
memory process of the volatility cluster phenomena by using
Minkowski metric for GARCH(1,1). Also we presented the result of the 
minus sign of the volatility in the reversed direction of the timescale.
It is named as the dark volatility or the hidden risk fear field.
These situations have been extensively studied and the correlations
have been found to be a very powerful tool.
Yet most natural
processes in the financial market are the non-stationary. In particular, in times of the financial crisis,
some accident events or economic shock news, stationarity is lost. \textbf{The precise definition of a nonstationary state of time series can help to analyze the non-equilibrium state in the financial market.}
As an example, we may
think about the financial market state from a mathematical
point of view. There is no any classical stochastic process
which will match with the real financial data, because there is not a single Kolmogorov space describing the whole
financial market. 
\textbf{ Dynamics of the market panic (in the financial world) from the point of view of new financial model is presented.  We hope that the precise
definition of Kolmogorov topological space can hopefully introduce
the better understanding of the macroeconomic models. The suitable algebraic reconstruction of space of time series could help with analysis of the prior effect and endeffect
of time series models. Our presented approach together with the close connection with of empirical mode decomposition and intrinsic time scale
decomposition of correlation matrix can serve as the main tools for a detection of
market crash over time series data, as was introduced in \cite{cohomo}.}

\section*{Acknowledgments}
\noindent

The work was partly supported by VEGA Grant No. 2/0009/16.
R. Pincak would like to thank the
TH division in CERN for hospitality. K. Kanjamapornkul  thanks to  100th Anniversary Chulalongkorn University Fund for Doctoral Scholarship.

\end{document}